\documentclass[aps,pre,longbibliography,twocolumn,floatfix]{revtex4-1}
\usepackage{graphicx}
\usepackage{amsmath}
\usepackage{amssymb}
\usepackage{bm}
\usepackage{times}
\usepackage[pdftex]{hyperref}
\hypersetup{colorlinks=true,linkcolor=blue,citecolor=blue,urlcolor=blue}
\usepackage{verbatim}

\begin{document}

\title{The droplet-scaling versus replica symmetry breaking debate in spin glasses revisited}

\author{M.~A.~Moore}
\affiliation{Department  of
Physics and Astronomy, University of Manchester, Manchester M13 9PL, United Kingdom}
\date{\today}

\begin{abstract}
Simulational studies of spin glasses in the last decade have focussed on the so-called replicon exponent $\alpha$ as a means of determining whether the low-temperature phase of spin glasses is described by the replica symmetry breaking picture of Parisi or by the droplet-scaling picture. On the latter picture, it should be zero, but we shall argue that it will only be zero for systems of linear dimension $L > L^*$. The crossover length $L^*$ may be of the order of hundreds of lattice spacings in three dimensions and approach infinity in 6 dimensions. We use the droplet-scaling picture to show that the apparent non-zero value of $\alpha$  when  $L < L^*$ should be $2 \theta$, where $\theta$ is the domain wall energy scaling exponent,  This formula is in reasonable agreement with the reported values of $\alpha$.
\end{abstract}
\maketitle

\section{Introduction}

The nature of the low temperatures phase of Ising spin glasses in finite dimensional spin glasses has been controversial for decades. A nice review of the situation was given by Newman and Stein in 2003 \cite{newman:03}. The two descriptions which are the most developed are the replica symmetry breaking picture (RSB) which derives from Parisi's exact solution \cite{parisi:79, parisi:83, rammal:86, mezard:87, parisi:08} of the Sherrington-Kirkpatrick model \cite{sherrington:75} and the droplet-scaling picture \cite{mcmillan:84a,bray:86, fisher:88}. There are two other pictures, the TNT picture of Krzakala and Martin \cite{krzakala:00} and of Palassini and Young \cite{palassini:00} and the chaotic pairs picture of Newman and Stein \cite{newman:98}. These four different pictures can be most readily distinguished by the nature of excitations or droplets produced from their ground state and the nature of the interfaces of the droplets or domain walls. Thus in $d$ dimensions consider the interface generated through changing the boundary conditions from periodic to anti-periodic in one direction in a cube of length $L$. The number of bonds in the interface will scale as $L^{d_s}$. If $d_s= d$ the interface is said to be space filling. In the RSB and chaotic pairs picture, interfaces are space filling. In the droplet-scaling and TNT picture the fractal dimension $d_s < d$. The other distinguishing feature of the four pictures is the (free) energy of the interfaces or droplets. In an Ising ferromagnet the energy of a domain wall separating \lq up' spins from \lq down' spins scales as $L^{d-1}$. In the droplet picture (and also the chaotic pairs picture) the energy of a spin glass interface or droplet is similar, increasing as $L^{\theta}$ and $\theta > 0$ when there is a finite temperature spin glass phase. However, it is different in the RSB and TNT picture. There an excitation or droplet can have an energy $O(1)$ even when it contains $O(L^d)$ spins.

It is my belief that what is the correct picture may change  with  the dimensionality $d$ of the system. The strong coupling renormalization group  
has been used \cite{monthus:15,wang:17,wang:18b} to study the value of $d_s$ as a function of dimensionality $d$. It was found that $d_s$ became equal to $d$ in six dimensions. This suggests that for dimensions $d > 6$ either the RSB picture or chaotic pairs could apply, while for $d < 6$ the droplet-scaling or TNT picture could apply.

Back in 2000 the TNT picture seemed to provide the description of the spin glass state which was best supported by simulational work in $d =3$. Simulations have mostly  been done on the Edwards-Anderson Ising spin Hamiltonian \cite{edwards:75} where the bonds $J_{ij}$ are between nearest-neighbors:
\begin{equation} 
\mathcal{H}=-\sum_{<ij>} J_{ij} S_i S_j-h\sum_i S_i.
\label{EAdef}
\end{equation}
A focus of many studies has been the Parisi overlap function \cite{parisi:79,parisi:83,parisi:08} between spins in two copies, $a$ and $b$ of the system, defined by 
\begin{equation}
P(q)= \overline{\langle\delta(q-\frac{1}{N}\sum_i S_i^a S_i^b)\rangle},
\label{Pqdef}
\end{equation} 
where the overline denotes the bond-average over the couplings $J_{ij}$. When the field $h= 0$,  $P(q)$ takes the trivial form of two delta functions in the droplet-scaling and chaotic pairs pictures, (at least in the thermodynamic limit when the number of sites $N\to \infty$);
\begin{equation}
P(q)=(1/2)\delta(q-q_{EA})+(1/2)\delta(q+q_{EA}),
\label{pqtriv}
\end{equation}
where $q_{EA}=(1/N)\sum_i \langle S_i\rangle^2$, calculated in the limit $h \to 0$. In the RSB and TNT pictures $P(q)$ is non-zero in the interval $-q_{EA} < q < q_{EA}$. 
Studies of $P(q)$ at, say,  $q =0$  showed that it remains finite as $L$, the linear dimension of the simulational box, is increased. However, in the droplet scaling picture it is predicted that $P(0)$ should decrease with $L$ at finite temperatures $T$, as $T/L^{\theta}$.  No  simulational study has ever seen any significant decrease of $P(0)$ with $L$ \citep{Banos:2010,Yucesoy:12,Billoire:13,Yucesoy:13}. On the other hand the study of interfaces seems to strongly support the idea that they are not space filling as $d_s < d$ (although naturally this was disputed \cite{marinari:00}). The initials TNT refer to the fact that the behavior of the  interface is \textit{trivial}, that is, as predicted by droplet scaling, but that the overlap function $P(q)$ is \textit{non-trivial} as in the RSB picture of Parisi and not as given by the trivial droplet-scaling prediction of Eq. (\ref{pqtriv}). 

Supporters of the droplet-scaling picture like the author of this paper would explain away this failure to predict the observed form of $P(q)$ in simulations as a \textit{finite size effect}. Studies of $P(q)$ have been restricted by computational limitations to systems whose linear dimension $L$ are usually less than 30. It is postulated that there is a length scale, $L^*$, which has to be surpassed before the true asymptotic behavior as $P(0) \sim T/L^{\theta}$ reveals itself.  Evidence that this  
might be a possibility has come from studies on the $d=2$ spin glass problem, which does not have a finite temperature spin glass phase ($\theta < 0$) but it has features which seem to have their analogue in $d =3$. In $d =3$ Krzakala and Martin \cite{krzakala:00} noted that there were excitations on the scale of their system size $L$ involving $O(L^d)$ spins whose energies were not as large as $~L^{\theta}$ but were instead of $O(1)$. Domain walls in $d = 3$ do have energies of $O(L^{\theta})$, but droplets seemed to exist which were as large as the system (in fact they often touched the boundaries of the system) but were of lower energy. In the droplet-scaling picture one considers a compact, connected  cluster of $N$ spins, of linear dimension $L$, such that $L^d < N < (2L)^d$, containing the chosen spin. It is assumed \cite{fisher:88, bray:86} that  the distribution  $\rho_L(E_L)$ of minimal energy clusters i.e. excitations has the scaling 
form 
\begin{equation}
\rho_L(E_L) \approx \frac{1}{\Upsilon L^{\theta}}\tilde{\rho}\bigg[\frac{E_L}{\Upsilon L^{\theta}}\bigg],
\label{dropletdist}
\end{equation} 
where $\Upsilon$ is a constant of the order of the standard deviation of the bonds $J_{ij}$ and $\tilde{\rho}(0) > 0$ for $d\ge 2$. (In Appendix A we shall calculate this distribution function analytically for the case of $d = 1$).  Eq. (\ref{dropletdist}) implies that the typical minimal droplet should have energy of $O(L^{\theta})$ and that the probability that the  minimal energy droplet has energy of $O(1)$ should fall off as $1/\Upsilon L^{\theta}$. It is this which lies behind the droplet-scaling prediction that $P(0)$ should decrease at temperature $T$ as $\sim T/\Upsilon L^{\theta}$. In fact in the study  of Ref. \cite{krzakala:00} there seemed to be more low energy droplets than expected from this formula  and it is this which is the basis of the TNT picture. A supporter of the droplet-scaling picture has to assert that the systems studied always have a size $L < L^*$.

This seems plausible if one looks at the behavior of droplets in $d = 2$ dimensions. The great advantage of studying two dimensions is that there exist polynomial time algorithms which enable one to obtain the ground state of very large systems. Thus by studying systems of size up to $10 000 \times 10 000$ it has been found that \cite{weigel:18} the energy associated with the change from periodic to anti-periodic boundary conditions has $\theta = -0.2793(3)$. The associated domain wall has a fractal dimension $d_s= 1.27319(9)$. However, the situation with droplets is  more complicated and produced a situation not unlike the debate between the advocates of the droplet picture and the RSB picture in three dimensions. The droplet scaling picture predicts that the correlation length, (as determined from the spin glass susceptibility), should grows as the temperature is reduced as $\xi(T) \sim 1/T^{1/|\theta|}$ \cite{bray:86} but the simulations at finite temperature found that they appeared to  grow with an effective exponent $\tilde{\theta} \sim -0.48$ \cite{Kawashima:92} down to the lowest temperatures  they could simulate. The origin of this discrepancy produced much controversy \cite{Kawashima:99,Berthier:03,
HartmannYoung:02,Picco:03}, before the correct explanation of the puzzle emerged \cite{HartmannMoorePRL,HartmannMoorePRB}. The key to its understanding lies in the fact that the droplet-scaling picture is indeed a scaling picture and that there are always corrections to the leading terms. These can be much larger in some quantities than in others. For example consider the domain wall produced by  changing the boundary conditions in one direction from periodic to antiperiodic. The standard deviation of the energy difference takes the form
\begin{equation}
\Delta E= AL^{\theta}+ BL^{-\omega},
\label{scalingcorr1}
\end{equation}
where the second term is the \lq \lq correction to scaling". Note that $\theta$ in two dimensions is negative (but is positive in three dimensions).  According to Ref. \cite{weigel:18} the correction to scaling term is very small for domain walls. The situation for droplets is very different and depends on how they are generated \cite{HartmannMoorePRL,HartmannMoorePRB}. Those which involve flipping the central spin but for which the resulting droplet did not touch the boundaries (the spins at the boundaries were fixed in their ground state orientations) e.g the cross droplets (see Ref. \cite{HartmannMoorePRL,HartmannMoorePRB} for details) are such that at small values of $L$, their energy seemed to decrease with an effective exponent $\tilde{\theta}=-0.47$ but for values of $L > L^* \approx 60$ a value close to the expected value of $-0.279$ was seen. Thus the simple droplet scaling behavior did emerge when droplets of large enough size could be studied. It would also be expected that the correlation length $\xi(T)$ would also grow according to droplet-scaling expectations if studied at low enough temperatures and this has now been confirmed by recent simulations \cite{Fernandez_2019}.

We turn now to three dimensional spin glasses. It is the contention of this paper that the finite energy droplets generated by the procedures used in Refs. \cite{krzakala:00,palassini:00} would not be of size $O(L^d)$ with a fractal dimension $d_s< d$ when $L>>L^*$. Here $L^*$ denotes the crossover length in \textit{three} dimensions, which I suspect might be even longer than its two-dimensional counterpart. Newman and Stein \cite{newman:01} have proved that in the large $L$ limit that excitations or droplets of size $O(L^d)$ with a fractal dimension $d_s< d$ cannot exist. (They proved  that the interfaces of these excitations must eventually pass  outside a fixed finite window, no matter how large, as the linear size L of the volume under consideration goes to
 infinity. So if TNT applied, then inside of any fixed window one
 eventually sees the same, single ground state pair (with, say, free or periodic
boundary conditions) in the large L limit, just as in the droplet-scaling picture).   The apparent evidence to the contrary in the numerical work of Refs.  \cite{krzakala:00,palassini:00} is because they were unable to simulate large enough systems and were working for system sizes $L < L^*$.  For systems larger than $L^*$ droplet-scaling features should emerge: If the finite energy droplets do not involve $O(L^d)$ spins when $L$ is large, they will not make the Parisi overlap function non-trivial and for $L > L^*$ the behavior in Eq.~(\ref{pqtriv}) will emerge.  Given that in two dimensions, the crossover length $L^* \approx 60$, then it seems likely  in three dimensions that the length scale $L^*$ will be even larger, perhaps of the order of hundreds of lattice spacings. I would anticipate that it will approach infinity as $d \to 6$ when these droplet states of $O(1)$ energy will have $d_s= d$ \cite{wang:17,wang:18b} and  produce  the RSB states expected for $d > 6$.

 On this viewpoint the TNT picture is just the droplet-scaling picture, with the recognition that there are  droplets whose energies are of $O(1)$ in systems whose linear dimensions are less than $L^*$ and  only  there do they have size $O(L^d)$.  This is all due to scaling corrections. In Ref. \cite{HartmannMoorePRL,HartmannMoorePRB} it was suggested that the possible  origin of these droplets whose energies are of $O(1)$ might arise from the fact that in Eq.~(\ref{scalingcorr1}) that $\Delta E$, now being used to describe the energy of droplets, has a minimum at some value of $L$ when $B > 0$ and $\theta> 0$. Around this minimum the $L$ dependence of $\Delta E$ will be small, giving rise to an effective value of $\theta$, $\tilde{\theta}$, which would be close to zero at the $L$ values near the minimum. It will be only at large $L> L^*$ that $\Delta E$ will clearly increase as $L^{\theta}$, just as occurred  for the cross-droplets in two dimensions when they were larger than $L^*$ \cite{HartmannMoorePRL,HartmannMoorePRB}. The size of corrections to scaling depend on the quantity being studied: Domain wall energies (for $\theta$) and the interface size (for the exponent $d_s$) may only have small corrections and the existing studies for $L< L^*$ may still be giving accurate answers for these exponents.
 
It would clearly be very desirable to have estimates of the value of $L^*$. This has been done by Middleton \cite{Middleton:13} for a particular form of the bond distribution, the $\pm J$ model. This bond distribution produces a macroscopic degeneracy for the ground state of the system and has  zero-energy droplets. The finite temperature excitations with free energies of $O(1)$ are discussed in \cite{Thomas:11}. At any non-zero temperature the properties of the $\pm J$ model should be similar to models with a continuous bond distribution \cite{Jorg:06} (but the value of $L^*$ will not be universal).  Middleton \cite{Middleton:13} estimated a value of $L^*$ in two dimensions of $\approx 64$. This is rather similar to the value $\approx 60$ obtained from the behavior of  the cross-droplets which were studied for  a Gaussian bond distribution in \cite{HartmannMoorePRL, HartmannMoorePRB}. One of his methods for getting a value of $L^*$ was similar to that used in Ref. \cite{Hatano:02} and using it Middleton obtained $L^*\approx 500$ in three dimensions.
 
Over the last decade the Janus collaboration and others  \cite{Janus:08, Banos6452,hartmann:15, billoire:17} have presented results which seem at first sight to be at variance with droplet-scaling expectations. They have found evidence which suggests that spin glasses in three dimensions have the behavior only expected of a system with RSB or chaotic pairs ordering. In Refs. \cite{Janus:08, hartmann:15} they carried out the following simulation. Starting from a randomly chosen set of spin configurations, they quenched to a temperature $T < T_c$, where $T_c$ is the spin glass transition temperature. They then let the spins evolve according to heat bath dynamics for a time $t_W$. This results in domains of spin glass order whose size is measured  by a coherence length $\xi(t_W)$ which grows as $t_W$ increases. They found that the  correlation function  
\begin{equation}
C_4(R_{ij},t_W) \equiv \overline{\langle S_i S_j\rangle^2} \sim \frac{1}{R_{ij}^{\alpha}} f\bigg(\frac{R_{ij}}{\xi(t_W)}\bigg).
\label{c4def}
\end{equation}
The overline is the usual bond average. In practice this was done by simulating two copies of the system with the same interaction but quenched into different initial random configurations, which allows an unbiased estimate of the thermal averages.   The function $f(x)$ falls off with increasing $x$ faster than exponentially and $f(x\to 0)={\rm const}$. The coherence length $\xi(t_W)$ is itself determined via the ratio of the second and zeroth moments of $C_4(R_{ij}, t_W)$. The
$k^{{\rm th}}$ moment is defined by
\begin{equation}
I_k(t_W)= \int d^d r\, r^k C_4(r,t_W),
\end{equation}
and then 
\begin{equation}
\xi(t_W)= \sqrt{\frac{I_2(t_W)}{I_0(t_W)}}.
\end{equation} 
The coherence length is found to grow slowly (coarsen) with $t_W$ as $\xi(t_W) \sim t_W^{{\rm const}/T}$. In simulations it can grow as large as $20$ to $ 30$ lattice spacings. Much interest attaches to
 the exponent $\alpha$ which is called the \lq \lq replicon exponent" by the Janus collaboration. An early estimate of its value in $d =3$ was $0.38(2)$ \cite{Janus:08}, while in a more recent paper this was revised downwards to $0.35 >\alpha >0.25$ \cite{Janus2018PRL}.
 
It seems likely that the coherence length $\xi(t_W)$ was always less than the crossover length $L^*$ in these simulations so one should expect TNT effects. Then the system will behave as if it has some RSB features (in particular, it will have droplets of size $O(L^d)$ with energy cost $O(1)$ which change as $\xi(t_W)$ grows). This will make the value of $\alpha$ appear to be non-zero.     
 The droplet-scaling prediction is \cite{hartmann:15,billoire:17},  $C_4(R_{ij},t_W)\to q_{EA}^2$  but this result will only be seen when $\xi(t_W)>> L^*$. This approach to a constant (which corresponds to $\alpha =0$) will emerge only in the limit $L^*<< R_{ij}<<\xi(t_W)$. The present simulations are a very long way off this limit. It means that the values for $\alpha$ currently being reported for $d =3$ are just effective values of this exponent, as they are only valid over a limited range of $R_{ij}$ and $\xi(t_W)$. 
 However, because $L^*$ may be quite large, the value for the replicon exponent $\alpha$ could be \textit{well-defined}: The crossover region where it gradually goes to its true value of zero has not been reached. In Sec. II  the values for $\alpha$ currently being reported are predicted by a simple argument which rests on the assumption that the correct picture of the three dimensional spin glass ordered phase is that of droplet-scaling.
We show that in dimensions $ d < 6$ that this effective value for $\alpha$ is $2 \theta$, where $\theta$ is the usual exponent describing the energy cost of a domain wall. This result is consistent with the numerical data on $\theta$ and $\alpha$, neither of which alas are very accurately determined at the present time. 

The exponent $\alpha$ appears in another form in studies of the metastates of spin glasses. Most metastates discussions in spin glasses concern equilibrium properties \cite{Newman:96,Newman:97,Newman:92,Newman:03b,Newman:15}. (One exception is Ref. \cite{fisher:06}). An exponent $\zeta$ has been introduced  and discussed at length by Read  \cite{read:14}. It is defined via the logarithm of the number of metastates which can be distinguished in a window of size $W$ which scales as $W^{d-\zeta}$. Read \cite{read:14} showed that the RSB picture predicted that $\zeta=4$ when $d >6$.  There is an assumed equivalence between the equilibrium metastates and those generated using a dynamical coarsening procedure to define an Aizenman-Wehr metastate \cite{aizenman:90}. If they are equivalent  $d-\zeta \equiv \alpha$, where $\alpha$ is defined from Eq.~(\ref{c4def}). In three dimensions Billoire et al. \cite{billoire:17} found using the equilibrium metastate approach  $\alpha=0.7 \pm 0.3$, while a coarsening dynamical metastate procedure was used by Manssen et al. who analysed their data with  $\alpha =0.438$ \cite{hartmann:15}. In Ref. \cite{Wittmann:16}  a simulation on a one-dimensional system with long-range interactions thought to be equivalent in its behavior to that of the EA model in $d = 8$ was used to construct the dynamical metastate and it gave a value for $\alpha$ consistent with Read's predictions and the assumed equivalence of static and dynamical metastate constructions.  Our expression for the effective exponent $\alpha$ agrees with Read's prediction at $d=6$.

In Sec. III we make suggestions for further (mostly simulational) work which could help in checking the validity of the scenario  advocated in this paper.

\section{The replicon exponent}
The Janus collaboration \cite{Janus2017,Janus2020,Janus2021}  also studied the effect of turning on a small field $h$ at time $t_W$  and determining the magnetization  $m(t+t_W)= (1/N)\sum_iS_i(t+t_W) $ at time $t+t_W$. They make the \lq\lq bold" claim \cite{Janus2017} that $m(t+t_W)$ has the form which one might have written down using equilibrium arguments, which they take to be
\begin{equation}
m(t+t_w,t_w;h)=\xi(t+t_W)^{y_h-d}\mathcal{F}(h [\xi(t+t_W)]^{y_h}; \mathcal{R}_{t,t_W}),
\label{fieldscaling}
\end{equation}
where $\mathcal{R}_{t,t_W} \equiv \xi(t+t_W)/\xi(t_W)$. The scaling function $\mathcal{F}(x,\mathcal{R})$ is odd in the $x$ argument for symmetry reasons. $t$ is of order $t_W$ so $\mathcal{R}_{t,t_W} \approx 1$ and dependence on this variable is ignored. The relevant length scale in this study is $\xi(t_W)$. The exponent $y_h$ was claimed to be related to $\alpha$ by
\begin{equation}
y_h= \frac{d}{2} - \frac{\alpha}{4}
\label{yhdef}
\end{equation}
We caution the reader at this point that the Janus collaboration have taken to writing what we call $\alpha$ as $\theta$. In this paper $\theta$ has its conventional spin glass meaning as the exponent associated with the domain or droplet energy (Janus call that $\zeta$).

 The magnetization can be written as a series expansion in odd powers of $h$:
\begin{equation}
m(h)=\chi_1 h+\frac{\chi_3}{3!}h^3+\frac{\chi_5}{5!}h^5 + O(h^7),
\label{series}
\end{equation}
where the dependency of the susceptibilities $\chi_1$, $\chi_3$, $\chi_5$ and $m$ on  $t$ and $t_W$ has been omitted to simplify the notation. Combining Eq. (\ref{fieldscaling}) and Eq. (\ref{series}), one deduces for example, that
\begin{equation}
\chi_3 \propto -\xi(t_W)^{4y_h -d}\equiv -\xi(t_W)^{d-\alpha}.
\label{xi3yh}
\end{equation}
In Ref. \cite{Janus2017} they wrote that \lq \lq At least in equilibrium $\chi_3$ is related to the space integral of the microscopic correlation function $C_4(R_{ij}, t_W)$". This if true would explain why $\alpha$ appears in $y_h$ as in Eq. (\ref{yhdef}). But it is only true when $T > T_c$. For $T < T_c$ the equilibrium expression for $\chi_3$ in terms of correlation functions was given long ago \cite{chalupa:77} and for a symmetric bond-distribution is  the bond-average of 
\begin{eqnarray}
\chi_3 &=& -\frac{6 \beta^3}{N}\sum_{ij} G_B(R_{ij}) \nonumber \\
&+&  \frac{4 \beta^3}{N} \sum _i (1-4 \langle S_i\rangle^2+3 \langle S_i\rangle ^4).
\label{chalupa}
\end{eqnarray}
where the correlation function $G_B(R_{ij})$ is the breather or  longitudinal correlation function studied in \cite{bray:79} and given by the bond average of
\begin{equation}
G_B(R_{ij})=\langle S_i S_j\rangle^2-4 \langle S_i\rangle \langle S_j\rangle \langle S_i S_j\rangle+ 3 \langle S_i \rangle^2 \langle S_j \rangle ^2.
\label{GBdef}
\end{equation}
The second line in Eq. (\ref{chalupa}) only contributes a finite term to $\chi_3$, whereas  the term involving $G_B$ gives a contribution which diverges with the size of the system. $G_B$ has been studied within the droplet-scaling picture \cite{bray:86, fisher:88}. There it was shown that it was given in terms of an integral involving the scaling function $\tilde{\rho}$ of Eq. (\ref{dropletdist}):
\begin{equation}
G_B(R_{ij}) \sim \frac{T q_{EA}^2}{\Upsilon R_{ij}^{\theta}}\int_0^{\infty} dx
\,\tilde{\rho}\bigg( \frac{ T x}{\Upsilon R_{ij}^{\theta}}\bigg) {\rm sech}^2 x(1-3 {\rm tanh}^2x).
\label{integral}
\end{equation}
For small values of the ratio $T/(\Upsilon R_{ij}^\theta)$ the integral can be approximated by setting the term in $\tilde{\rho}$ to its value at $\tilde{\rho}(0)$. However, the integral then is zero, so one has to expand $\tilde{\rho}(x)$ to next order in its Taylor series expansion; $\tilde{\rho}(x)= \tilde{\rho}(0) - A x+\cdots$. Then $G_B(R_{ij})$ becomes
\begin{equation}
G_B(R_{ij})\sim A q_{EA}^2 \bigg(\frac{T}{\Upsilon R_{ij}^{\theta}}\bigg)^2.
\label{GB}
\end{equation}
Previously \cite{bray:86,fisher:88}, out of an abundance of caution, the possibility that $\tilde{\rho}(x)=\tilde{\rho}(0)-A x ^{\phi}$ was considered. This changes the exponent of the term in $T/\Upsilon R_{ij}^{\theta}$ in Eq.~(\ref{GB}) from $2$ to $(1+\phi)$.  In Appendix A it is shown for the case of $d=1$ that the Taylor series form for $\tilde{\rho}(x)$ is appropriate. The numerical data in $d = 2$ \cite{HartmannMoorePRB} is at least consistent with the  Taylor series expansion form  $\phi=1$. We shall in this paper from now on just take $\phi = 1$. 

Using Eq.~(\ref{GB}) for $G_B(R_{ij})$ one deduces that $\chi_3 \sim - L^{d-2 \theta}$. Compare this with  Eq.~(\ref{xi3yh}); $\chi_3 \sim - \xi(t_W)^{d-\alpha}$ in the coarsening investigation. In coarsening the relaxation modes of the system with wavevectors greater than $1/\xi(t_W)$ are equilibrated. At wavevectors $k < 1/\xi(t_W)$ the system will still have the imprint of the infinite temperature system it was before the quench. This suggests that this region of $k$-space will only make a finite contribution to $\chi_3$. Hence we should be able to  equate the exponents of $L$ and $\xi(t_W)$ in the two expressions for $\chi_3$, so making
\begin{equation}
\alpha = 2 \theta.
\label{alphatheta}
\end{equation}
The equilibration of the system within the length scale $\xi(t_W)$ is not such as happens in a finite system on ergodic time scales which drives $\langle S_i \rangle$ to zero. Instead it is more like the equilibration of an infinite system in which boundary conditions are applied to break the up-down symmetry and leave $\langle S_i \rangle$ non-zero. It is to this situation that Eqs. (\ref{chalupa}) and (\ref{GBdef}) are applicable.

We next compare Eq.~(\ref{alphatheta}) with the current numerical estimates of $\alpha$ and $\theta$. Alas neither are known with great precision. In $d = 3$, $\alpha$ was found to be $0.38(2)$ in \cite{Janus:08}. In later work \cite{Janus2018PRL}, they noticed  there was an apparent temperature dependence in the value of $\alpha$. If the quench were not to a temperature less than $T_c$ but instead to $T_c$ itself $\alpha = d-2+\eta \approx 0.610(4)$ (see the definition of $C_4(R_{ij}, t_W)$ in Eq.~(\ref{c4def})). They argued that their apparent temperature dependence was due to proximity to $T_c$ in their work. This effect would go away if $\xi(t_W)\to \infty$. However, by using an extrapolation  to this limit they estimated  that $0.35 >\alpha > 0.25$. The value for $\theta$ was given  by Hartmann \cite{hartmann3d:99} to be $0.19(2)$ and by Boettcher \cite{boettcher:05} as $0.24(1)$. For $d =4$ both exponents are even less precisely determined: $\alpha=1.03(2)$ according to Ref. \cite{parisi4D:14}, while $\theta$ is $0.64(5)$ according to Ref. \cite{hartmann4d:99} and $0.61(1)$ according to Ref. \cite{boettcher:05}. In $d =6$, the RSB formula of Read \cite{read:14} 
 gives $\alpha = 2$, while the RSB based calculations of Ref. \cite{aspelmeier:16b} give $\theta = 1$. The numerical estimate in $d= 6$ by Boettcher in \cite{boettcher:05} was $\theta=1.1(1)$. I would judge that the agreement of Eq.~(\ref{alphatheta}) with the data for $d \le 6$ to be satisfactory, given the large uncertainties in the numerical values of $\alpha$ and $\theta$.
 
\section{Discussion}
In this paper we have argued that the old TNT picture of spin glasses can explain not only the old puzzle of the behavior of the Parisi overlap function $P(q)$ but also the more recent results of the Janus group on the correlation function $C_4(R_{ij},t_W)$. It has also been argued that the TNT picture is only relevant when phenomena on length scales $< L^*$ are studied, and that the RSB-like behavior seen on length scales less than $L^*$ will change to droplet-scaling behavior on the longest length scales. It has been possible to show the replicon exponent $\alpha$ of the Janus collaboration is equal to $2 \theta$ for $d <6$. This result shows that even in the RSB-like region $L <L^*$, droplet-scaling calculations have utility.

We suspect that the length scale $L^*$ in $d = 3$ might be so large that any crossover in behavior might be impossible to see in simulations at present, where $\xi(t_W)$ or the linear size $L$ of an equilibrated system are usually less than $30$. However, in experiments it has been claimed \cite{zhai:19} that  values of $\xi(t_W)$ of order $250$  are being seen. In $d = 2$ a crossover was seen at values of $L^*=60$ \cite{HartmannMoorePRB} which suggests that $L^*$ could be hundreds  of lattice spacings for $d=3$. Indeed  $L^*$ probably grows to infinity as $d \to 6$ when the droplets of energy $O(1)$ of the TNT picture become the pure states of RSB. Our value for $\alpha$ coincides  with that derived from the  RSB picture by Read \cite{read:14} in six dimensions. However, the large value of $L^*$ allow the possibility that for sizes $L < L^*$ the apparent values quoted for $\alpha$ could be well-defined and not influenced by the expected  creep towards $0$ when $L$ grows past $L^*$.

I shall now make a few suggestions for a number of investigations which might help to clarify what is going on. 
\begin{enumerate}
\item If indeed $L^*$ is of order $500$ in three dimensions  many properties of spin glasses will appear both in simulations and experiments to be just as expected from the RSB picture. However, a key difference between RSB (and chaotic pairs) and droplet-scaling is that on the droplet-scaling picture $d_s < d$, whereas on the other two pictures, the domain wall produced by (for example) a change of boundary conditions from periodic to anti-periodic in one direction, is space filling with $d_s = d$. Domain wall energies seem to have much smaller corrections to scaling than those of droplets and so studying domain walls would seem likely to be a good way of also getting at $d_s$. Much numerical work, although it was done for $L <L^*$, does favor $d_s < d$, thus supporting the droplet-scaling picture. Exponents like $\theta$ and $d_s$ are exponents associated with the zero-temperature fixed point and it would be natural to expect that the best results for their value would be obtained from  $T=0$ studies.  Alas, that requires finding ground states of the Hamiltonian which is NP hard for $d> 2$. The methods which have been used at finite temperature usually involve extensive data manipulation \cite{Janus:08}. An old simulation by Huse \cite{huse:91} gave a way of determining $d_s$ at finite temperatures using a coarsening procedure which avoided extensive data manipulation (and yielded $d_s < d$). That approach could nowadays be pushed to larger values of $\xi(t_W)$.

\item The susceptibility $\chi_3$ is an integral over all space of the correlation function $G_B(R_{ij})$. This correlation function was studied using simulations in $d=3$ in Ref. \cite{Giardina:05} but only for rather modest system sizes  ($L = 12$). Both   RSB and droplet scaling predict a power law  decrease of $G_B$ with $R_{ij}$. In Ref. \cite{Giardina:05} a more rapid decay, possibly exponential, with $R_{ij}$, was seen. That might be due to finite size effects, but it would be useful if this topic could now be re-visited.

\item The  predictions of the Janus collaboration of the behavior of $\chi_3$ as $L^{d-\alpha}$ should presumably extend to the  RSB region $d >6$. Using Read's result $\alpha=d-4$ for $d >6$, the divergence of $\chi_3$ is then  $\chi_3 \sim L^4$. The analytical work in Ref. \cite{Giardina:05} predicts a divergence of $\chi_3$ from $G_B$ as $L$ in dimensions $d >6$. It would be interesting to study this discrepancy using the one-dimensional proxy model for high dimensions used in Ref. \cite{Wittmann:16}. However, ageing a system with RSB towards equilibrium needs to be re-examined in the light of the recent findings of \cite{Bernaschi:20} for the Viana-Bray model \cite{Viana_1985} who found that the system stayed trapped in a confined region of the configuration space.
 \end{enumerate} 

Within the spin glass phase itself the large value of the crossover length $L^*$ will make it difficult to provide good numerical or experimental evidence as to which picture of spin glasses, droplet-scaling or RSB, is correct. $L^*$ is, however, a feature of the zero-temperature fixed point. Fortunately there is another way to resolve the debate which avoids features produced by the zero-temperature fixed point and that is to determine whether there is an de Almeida-Thouless  transition \cite{dealmeida:78} when a magnetic field is applied. This line marks the onset  to a state with RSB, on cooling in a field and is absent according to the droplet-scaling picture \cite{bray:86,fisher:88}. There have been doubts as to its existence below six dimensions ever since Bray and Roberts \cite{bray:80} were unable to find a stable perturbative critical fixed point for it in dimensions $d$ just below $6$. Further arguments to this effect have been given \cite{Moore:11, moore:14}. In three dimensions there is experimental evidence \cite{nordblad:95} supporting the absence of the de Almeida-Thouless line. Simulations on this issue \cite{larson:13} provide in the view of this author excellent evidence that there is no de Almeida-Thouless transition, (but some still remain of the view that there is a transition \cite{Janus2021}; finite size complications \cite{yeo:15, larson:16} are not insignificant).

\begin{acknowledgments}
I should like to acknowledge very useful exchanges with Alex Hartmann, Chuck Newman, Daniel Stein, Victor Martin-Mayor, Nick Read, and Peter Young.
\end{acknowledgments}

\appendix*
\section {The droplet distribution function in one dimension}
\label{appendix}
In Sec. II we made the assumption that the scaling function of Eq.~(\ref{dropletdist}), $\tilde{\rho}(x)$, had a Taylor series expansion as $\tilde{\rho}(x)= \tilde{\rho}(0)- A x+ \cdots$ rather than as  $\tilde{\rho}(x)= \tilde{\rho}(0)- A x^{\phi}$. In this Appendix the scaling function is obtained analytically for $d = 1$ and it is shown that in this case $\phi = 1$ and that the Taylor series expansion of $\tilde{\rho}(x)$ is valid.

\begin{figure}
  \includegraphics[width=\columnwidth]{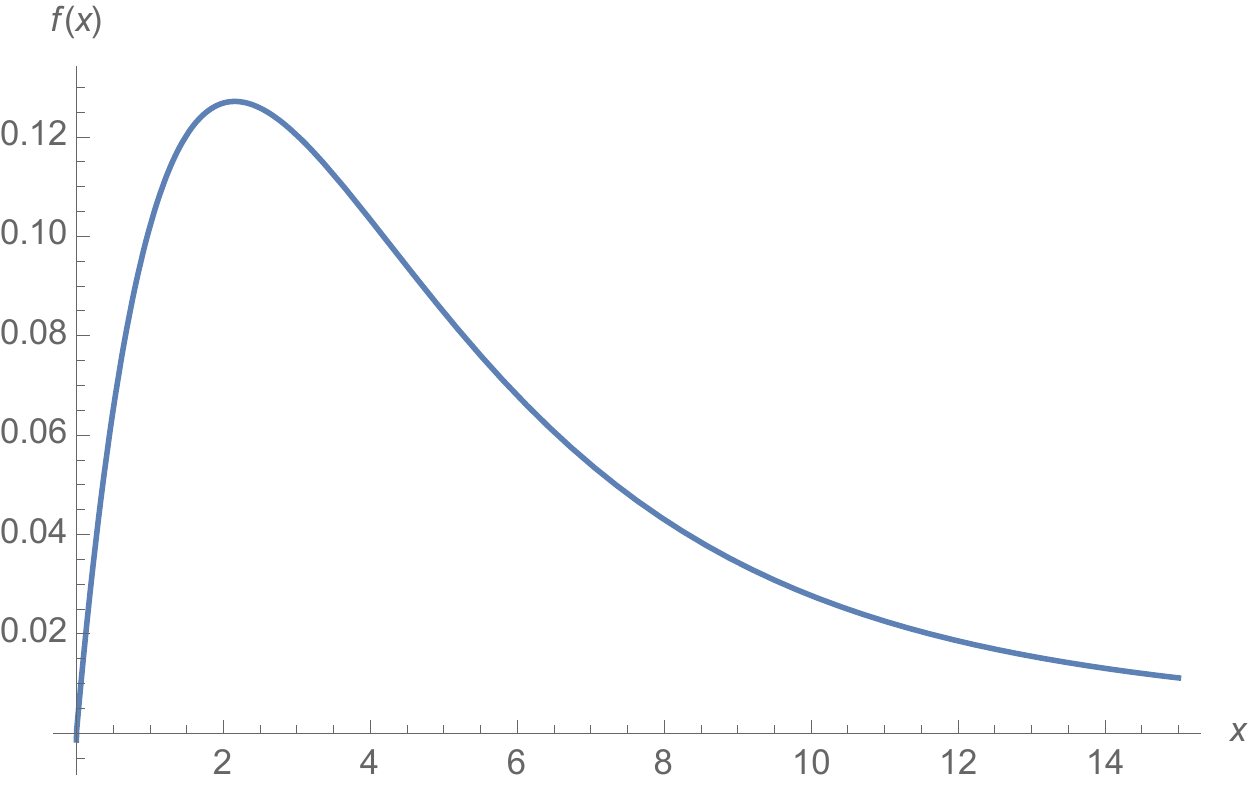}
  \caption{The scaling function $f(x)=\tilde{\rho}(x)$ for the droplet energy distribution scaling function in one dimension.}
  \label{fig: scalingfunc}
\end{figure}

The ground state of a one-dimensional spin system with open boundary conditions is found by making (say) the spin at one end, $S_1=1$ and fixing the orientation of the remaining spins using $S_i S_{i+1}=sign(J_{i, i+1})$. To find the domain wall energy one flips the spin $S_N$. This causes all the spins up to the bond of smallest magnitude to flip. In a system of $L$ bonds, the bond $|J|$ of smallest magnitude  has the distribution for large $L$ \cite{bray:86}
\begin{equation} 
P_L(|J|)= \frac{1}{J(L)}\exp\bigg(-\frac{|J|}{J(L)}\bigg),
\label{domdist}
\end{equation}
 and 
\begin{equation}
J(L)=\frac{1}{P_J(0)L}.
\label{theta1d}
\end{equation}
 Here $P_J(0)$ denotes the value of the bond distribution function at $J= 0$, and provided it is non-zero,  Eq.~(\ref{theta1d}) implies that $\theta = -1$ for $d =1$. Eq.~(\ref{domdist}) is the distribution function for  \lq \lq domain wall energies" $E$, which is  equal to that of $|J|$.

The minimal droplet energy $E$ around site $i$ is the sum of the energy of the  bond, $E_1$, to the right of site $i$ where all the bonds between $i$ and the bond to the right of the site at $i+L_1$ have magnitudes greater than $E_1=|J_{i+L_1, i+L_1+1}|$ plus the energy of the  bond, $E_2$, to the left of site $i$ where all the bonds between $i$ and the bond to the left of the site at $i-L_2$ have magnitudes larger than $E_2=|J_{i-L_2, i-L_2-1}|$. Then the $L=L_1+L_2$ spins lying between $i-L_2$ and $i+L1$ can all be flipped together at a total energy cost of $E=E_1+E_2$.  For large $L_1$ and $L_2$ the distribution of $E_1$ and $E_2$ will be as given in Eq.~(\ref{domdist}), so the probability distribution of droplets of energy $E$ and size $L$ will be
\begin{multline} 
\rho_L(E)=\frac{1}{L}\int dL_1\int dL_2 \int dE_1\int dE_2  \\ \delta (E-E1-E_2)\delta(L-L_1-L_2)P_{L_1}(E_1)P_{L_2}(E_2).
\end{multline}
The coefficient $1/L$ arises as the spin i can be in any of the $L$ sites between the weak bonds which are broken when the droplet is flipped. Writing $\rho_L(E)$ in terms of its scaling form $\rho_L(E)=(1/J(L)) \tilde{\rho}(E/J(L))$ one finds 
\begin{eqnarray}
\tilde{\rho}(x)&=& \int_0^1 \,dy  \frac{y (1-y) [\exp(-xy)-\exp(-x(1-y))]}{1-2 y} \nonumber \\
&=&\frac{1}{2}e^{-x}\bigg(-\frac{2+e^{x}(-2+x)+x}{x^2}+e^{x/2}{\rm Shi(x/2)}\bigg),\nonumber
\end{eqnarray}
where ${\rm Shi}$ is the $\sinh$ integral and $x=E/J(L)$. The function $\tilde{\rho}(x)$ in shown in Fig. 1.

In two dimensions $\tilde{\rho}(0)$ is finite rather than as here zero. The function $\tilde{\rho}(x)$ is an increasing function of $x$ at small values of $x$ just as in two dimensions \cite{HartmannMoorePRB}. In $d = 3$ I suspect that $\tilde{\rho}(x)$ is actually a decreasing function of $x$. Fig. \ref{fig: scalingfunc} shows that the scaling function has a long tail at large $x$; in fact it is so long-tailed that the mean droplet size is not well-defined. $\tilde{\rho}(x)$ has the Taylor series expansion at small $x$
\begin{equation}
\tilde{\rho}(x)=x/6-x^2/12 +x^3/45 -O(x)^4,
\end{equation}
which implies that $\phi = 1$ in one dimension.

\bibliography{refs}
\end{document}